\documentclass[journal]{IEEEtran}
\usepackage{ifpdf}
\usepackage{booktabs}

\usepackage{cite}

\ifCLASSINFOpdf
\usepackage[pdftex]{graphicx}
\else
\usepackage[dvips]{graphicx}
\fi
\usepackage{amsmath}
\usepackage{algorithmic}

\usepackage{array}

\ifCLASSOPTIONcompsoc
\usepackage[caption=false,font=normalsize,labelfont=sf,textfont=sf]{subfig}
\else
\usepackage[caption=false,font=footnotesize]{subfig}
\fi
\usepackage{fixltx2e}

\usepackage{stfloats}
\ifCLASSOPTIONcaptionsoff
\usepackage[nomarkers]{endfloat}
\let\MYoriglatexcaption\caption
\renewcommand{\caption}[2][\relax]{\MYoriglatexcaption[#2]{#2}}
\fi
\usepackage{url}

\usepackage{bm}
\usepackage{enumitem}
\usepackage{color}
\renewcommand{\figurename}{Fig. }
\renewcommand{\tablename}{Tab. }
\usepackage{multirow}
\usepackage{cite}
\usepackage{diagbox}

\begin{document}
	%
	\title{Network Algebraization and Port Relationship for Power-Electronic-Dominated Power Systems}
	%
	%
	%

\author{Rui~Ma,
		Xiaowen~Yang,
		Meng~Zhan,~\IEEEmembership{Senior Member,~IEEE}
		\thanks{
		R. Ma, X. Yang and M. Zhan are with the State Key Laboratory of Advanced Electromagnetic Engineering and Technology, School of Electrical and Electronic Engineering, Huazhong University of Science and Technology, Wuhan 430074, China (e-mail: d201880429@hust.edu.cn; m202071528@hust.edu.cn; zhanmeng@hust.edu.cn).
		}
}
	\maketitle

\begin{abstract}
Different from the quasi-static network in the traditional power system, the dynamic network in the power-electronic-dominated power system should be considered due to rapid response of converters' controls.
In this paper, a nonlinear differential-algebraic model framework is established with algebraic equations for dynamic electrical networks and differential equations for the (source) nodes, by generalizing the Kron reduction. 
The internal and terminal voltages of source nodes including converters are chosen as ports of nodes and networks.
Correspondingly, the impact of dynamic network becomes clear, namely, it serves as a voltage divider and generates the terminal voltage based on the internal voltage of the sources instantaneously, even when the dynamics of inductance are included.
With this simplest model, the roles of both nodes and the network become apparent.
Simulations verify the proposed model framework in the modified 9-bus system.
\end{abstract}


\section{Introduction}
\IEEEPARstart{W}{ith} wide application of renewable energy resources via power electronic devices, an increasing number of synchronous generators (SGs) have been replaced by inverter-based resources. This has substantially changed the dynamics of power-electronic-dominated power system (PEDPS) \cite{schiffer2016survey,ji2020dynamic,2021definition}.
In the SG-dominated traditional power system, it has been well recognized that the system electromechanical dynamics is described by differential-algebraic equations where the node for the SG (including the swing motion and internal voltage dynamics) is depicted by differential equations and the network by a nodal admittance matrix (algebraic equations) of phasors \cite{kundur1994power}. 
Correspondingly, due to the rapid response of converters' controls in the PEDPS, the dynamics of loads and transmission lines should be considered, especially for high-frequency oscillation analysis \cite{green2007modeling}. 
It is suggested that whole-system, differential, vector, electromagnetic-transient (EMT) simulation should be needed.

The state-space model, impedance model and harmonic state-space model of the line inductance were established to study sub- and super-synchronous oscillations \cite{chi2019overview}.
In addition, it was found that the quasi-static network model for low-frequency oscillations ($<$10Hz) was applicable, while dynamics of networks should be considered for higher-frequency oscillations \cite{yang2022dynamic}.
Meanwhile, the impacts of quasi-static and dynamic phasors on the system stability were systematically evaluated very recently \cite{vega2021analysis}. Within the DC-bus control timescale, differential-algebraic equations of multi-converter systems were established by neglecting all electromagnetic fast transients \cite{yang2020multi}.
\emph{Nevertheless, the network dynamic characteristics and its role in the PEDPS remains unclear, to the best of our knowledge.}

In this paper, inspired by results in the single machine scenario, an algebraic description of the dynamic electrical network is established, even when the dynamics of inductance are included. Then, the property of networks is also revealed based on the proposed differential-algebraic modeling framework. Finally, the proposed model is verified through simulations in the modified IEEE 3-machine-9-bus system.

\section{Single VSC System}
\begin{figure}[htbp!]
	\centering
	\includegraphics[width=0.9\linewidth]{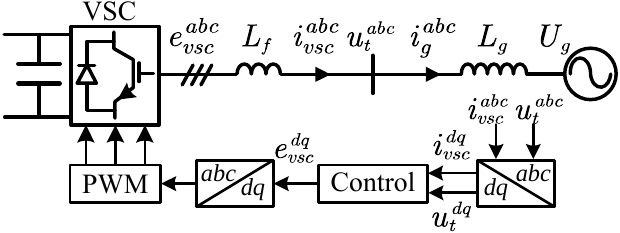}
	\caption{Single VSC system with typical controls.}
	\label{topology}
\end{figure}

Let us start from a single voltage source converter (VSC) system \cite{ma2020sustained}, which is connected to the AC grid ($U_g$) by a filter inductance $L_{f}$ and a grid inductance $L_{g}$ in \figurename \ref{topology}. It usually generates the internal voltage $e_{vsc}^{abc}$ (defined as the fundamental component of the PWM output) by the alternative current control (ACC), terminal voltage control, and phase-locked loop (PLL). The variables in the common ($xy$) and local ($dq$) synchronous reference frames can be transferred to each other by coordinate transformation.

The state equations of currents on the filter inductance ($i_{vsc}^{xy}$) are
\begin{equation}
\label{network1_d}
\left\{ \begin{array}{l}
\dot i_{vsc}^x = \frac{{{\omega _0}}}{{{L_f}}}e_{vsc}^x - \frac{{{\omega _0}}}{{{L_f}}}u_t^x + {\omega _0}i_{vsc}^y,\\
\dot i_{vsc}^y = \frac{{{\omega _0}}}{{{L_f}}}e_{vsc}^y - \frac{{{\omega _0}}}{{{L_f}}}u_t^y - {\omega _0}i_{vsc}^x,
\end{array} \right.
\end{equation}
and those of currents on the grid inductance ($i_{g}^{xy}$) are
\begin{equation}
\left\{ \begin{array}{l}
\dot i_g^x = \frac{{{\omega _0}}}{{{L_g}}}u_t^x - \frac{{{\omega _0}}}{{{L_g}}}u_g^x + {\omega _0}i_g^y,\\
\dot i_g^y = \frac{{{\omega _0}}}{{{L_g}}}u_t^y - \frac{{{\omega _0}}}{{{L_g}}}u_g^y - {\omega _0}i_g^x.
\end{array} \right.
\end{equation}
As $i_{vsc}^{xy}= i_{g}^{xy}$, one of them is superfluous according to circuit theory.
Consequently, considering (1) and (2), we have the terminal voltage ($u_{t}^{xy}$) as a function of the internal voltage ($e_{vsc}^{xy}$) and the grid voltage ($u_{g}^{xy}$) \cite{ma2020sustained}, i.e.,
\begin{equation}
\left\{ \begin{array}{l}
u_t^x = \frac{{{L_g}}}{{{L_f} + {L_g}}}e_{vsc}^x + \frac{{{L_f}}}{{{L_f} + {L_g}}}u_g^x, \\
u_t^y = \frac{{{L_g}}}{{{L_f} + {L_g}}}e_{vsc}^y + \frac{{{L_f}}}{{{L_f} + {L_g}}}u_g^y.
\end{array} \right.
\label{network1}
\end{equation}

\begin{figure}[htbp]
	\centering
	\includegraphics[width=0.8\linewidth]{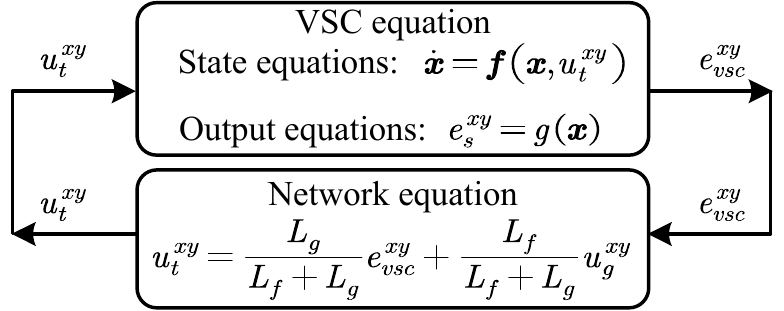}
	\caption{Schematic show for the single VSC system.}
	\label{single_VSC}
\end{figure}

Therefore, the dynamic network can be described by the linear algebraic equations (\ref{network1}) and the filter inductance dynamics are incorporated into the differential equations of the VSC in (\ref{network1_d}). The system is re-structured in \figurename \ref{single_VSC}, where the internal and terminal voltages are chosen as the output and input of the VSC, respectively, and accordingly, they are the input and output of the network. Hence, only independent state variables are kept, accompanying with an already algebraized network.
Below let us see the general form in the multi-machine system.

\section{Multi-Machine System}
For a PEDPS, it usually includes VSC, resistance-inductance ($RL$) loads, an infinite bus, and interconnecting transmission lines. Assume that the line resistance and ground capacitance are neglected. 
First, we consider the branch current dynamics from the $i$th bus to the $j$th bus ($i_{ij}^{xy}$):
\begin{equation}
\label{node_current_1}
\left\{
\begin{array}{l}
\dot i_{ij}^x = \frac{{{\omega _0}}}{{{L_{ij}}}}(u_{ti}^x - u_{tj}^x) + {\omega _0}i_{ij}^y,\\
\dot i_{ij}^y = \frac{{{\omega _0}}}{{{L_{ij}}}}(u_{ti}^y - u_{tj}^y) - {\omega _0}i_{ij}^x,
\end{array}
\right.
\end{equation}
where $u_{ti}^{xy}$ and $u_{tj}^{xy}$ represent the $i$th and $j$th bus voltages, respectively.
$L_{ij}$ ($i \neq j$) is the mutual inductance between nodes $i$ and $j$, and $L_{ij}=0$ in the absence of a connection.
By summarizing all branch currents ending at the node $i$, from (\ref{node_current_1}) we obtain
\begin{equation}
\left\{ \begin{array}{l}
\dot i_i^x = {\omega _0}u_{ti}^x\sum\limits_{j = 1}^n {\frac{1}{{{L_{ij}}}}}  - {\omega _0}\sum\limits_{j = 1}^n {\frac{1}{{{L_{ij}}}}u_{tj}^x}  + {\omega _0}i_i^y , \\
\dot i_i^y = {\omega _0}u_{ti}^y\sum\limits_{j = 1}^n {\frac{1}{{{L_{ij}}}}}  - {\omega _0}\sum\limits_{j = 1}^n {\frac{1}{{{L_{ij}}}}u_{tj}^y}  - {\omega _0}i_i^x,
\end{array} \right.
\label{node_current}
\end{equation}
where the total current injections are $i_{i}^{x} = \sum\limits_{j = 1}^n i_{ij}^{x}$ and $i_{i}^{y} = \sum\limits_{j = 1}^n i_{ij}^{y}$.

\begin{figure}[htbp!]
	\centering
	\includegraphics[width=0.9\linewidth]{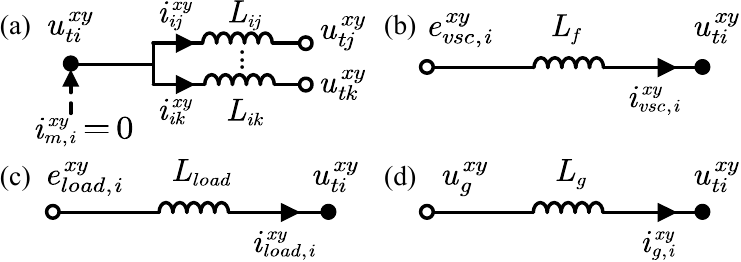}
	\caption{Relations between internal and terminal voltages for different types of node: (a) intermediate node, (b) VSC node, (c) load node, and (d) slack node. }
	\label{nodes}
\end{figure}

Next four typical nodes including the intermediate, VSC, load, and slack nodes in \figurename \ref{nodes} will be separately studied.

\subsubsection{Intermediate node}As the current injection equals zero in \figurename \ref{nodes}(a), i.e., $i_{m,i}^{xy} = 0$, from (\ref{node_current}) we have
\begin{equation}
{Y_{ii}}u_{ti}^x + \sum\limits_{j = 1}^n {{Y_{ij}}u_{tj}^x}  = 0,
\label{inter}
\end{equation}
where
\begin{equation}
{Y_{ii}} = \sum\limits_{j = 1}^n {\frac{1}{{{L_{ij}}}}}, ~ {Y_{ij}} =  - \frac{1}{{{L_{ij}}}}.
\end{equation}
Here $Y_{ii}$ and $Y_{ij}$ are self- and mutual- admittances, respectively. Similarly, we can obtain the corresponding equations for the $y$ axis.

\subsubsection{VSC node}The dynamics of current $i_{i}^{xy}$ of the filter inductance in \figurename \ref{nodes}(b) can be written as
\begin{equation}
\left\{ \begin{array}{l}
\dot i_{vsc,i}^x = \frac{{{\omega _0}}}{{{L_f}}}e_{vsc,i}^x - \frac{{{\omega _0}}}{{{L_f}}}u_{ti}^x + {\omega _0}i_{vsc,i}^y, \\
\dot i_{vsc,i}^y = \frac{{{\omega _0}}}{{{L_f}}}e_{vsc,i}^y - \frac{{{\omega _0}}}{{{L_f}}}u_{ti}^y - {\omega _0}i_{vsc,i}^x.
\end{array} \right.
\label{VSC_current}
\end{equation}
As this current equals the injected current at the VSC node, i.e., $i_{vsc,i}^{xy} = i_i^{xy}$, by considering (\ref{node_current}) and (\ref{VSC_current}) we have
\begin{equation}
{Y_{ii}}u_{ti}^x + \sum\limits_{j = 1}^n {{Y_{ij}}u_{tj}^x = \frac{1}{{{L_f}}}e_{vsc,i}^x},
\label{vsc_node}
\end{equation}
where similarly the self- and mutual- admittance for the VSC node, respectively, are
\begin{equation}
{Y_{ii}} = \frac{1}{{{L_f}}} + \sum\limits_{j = 1}^n {\frac{1}{{{L_{ij}}}}}, ~ ~ {Y_{ij}} =  - \frac{1}{{{L_{ij}}}}.
\end{equation}

\subsubsection{Load node}As shown in \figurename \ref{nodes}(c), similarly we have
\begin{equation}
{Y_{ii}}u_{ti}^x + \sum\limits_{j = 1}^n {{Y_{ij}}u_{tj}^x = \frac{1}{{{L_{load}}}}e_{load,i}^x},
\label{load_node}
\end{equation}
where
\begin{equation}
{Y_{ii}} = \frac{1}{{{L_{load}}}} + \sum\limits_{j = 1}^n {\frac{1}{{{L_{ij}}}}},
~{Y_{ij}} =  - \frac{1}{{{L_{ij}}}}, 
~e_{load,i}^x =  - {r_{load}}i_i^x.
\end{equation}
Here $L_{load}$ and $r_{load}$ denote the inductance and resistance of the load, respectively, and $e_{load,i}^{xy}$ is defined as the equivalent internal voltage of the load.

\subsubsection{Slack node}Finally, as shown in \figurename \ref{nodes}(d), we have
\begin{equation}
{Y_{ii}}u_{ti}^x + \sum\limits_{j = 1}^n {Y_{ij}} u_{tj}^x =  \frac{1}{{{L_g}}}u_g^x,
\label{grid_node}
\end{equation}
where
\begin{equation}
{Y_{ii}} = \frac{1}{{{L_{g}}}} + \sum\limits_{j = 1}^n {\frac{1}{{{L_{ij}}}}},~ ~ {Y_{ij}} =  - \frac{1}{{{L_{ij}}}}.
\end{equation}

In summary, after combining (\ref{inter}), (\ref{vsc_node}), (\ref{load_node}) and (\ref{grid_node}) for all these different types of node, the dynamic network can be uniformly described by a matrix,
\begin{equation}
{\bf{Yu}}_t^{xy} = {{\bf{Y}}_{fr}}{{\bf{e}}^{xy}},
\label{ori_network}
\end{equation}
where ${{\bf{Y}}}$ denotes the nodal admittance matrix and
${{\bf{Y}}_{fr}}$ is a diagonal matrix composed of $1 / L_f$ (VSC node), $1 / L_{load}$ (load node), $1 / L_g$ (slack node), and $0$ (intermediate node). Separating all nodes as source nodes (including the VSC, load, and infinite bus) and intermediate nodes, we have
\begin{equation}
\left[ {\begin{array}{*{20}{c}}
	{{{\bf{Y}}_a}}&{{{\bf{Y}}_b}}\\
	{{{\bf{Y}}_c}}&{{{\bf{Y}}_d}}
	\end{array}} \right]\left[ {\begin{array}{*{20}{c}}
	{{\bf{u}}_{t,s}^x}\\
	{{\bf{u}}_{t,m}^x}
	\end{array}} \right] = \left[ {\begin{array}{*{20}{c}}
	{{{\bf{Y}}_{f}}}&{\bf{0}}\\
	{\bf{0}}&{\bf{0}}
	\end{array}} \right]\left[ {\begin{array}{*{20}{c}}
	{{\bf{e}}_s^x}\\
	{\bf{0}}
	\end{array}} \right],
\label{ini_net}
\end{equation}
where ${\bf{Y}} = [{{\bf{Y}}_a},{{\bf{Y}}_b};{{\bf{Y}}_c},{{\bf{Y}}_d}]$.
${\bf{u}}^{x}_{t,s}$ and ${\bf{u}}^{x}_{t,m}$ represent the terminal voltages of the source and intermediate nodes, respectively. ${\bf{e}}^{x}_{s}$ represents the internal voltage of the sources.
${{\bf{Y}}_{f}}$ is a full diagonal matrix composed of $1 / L_f$, $1 / L_{load}$, and $1 / L_g$.
We can further eliminate all intermediate nodes, similar to the Kron reduction in the traditional power system analysis \cite{kundur1994power}, and have
\begin{equation}
{\bf{u}}_{t,s}^{xy} ={{\bf{Y}}^{-1}_r} {{\bf{Y}}_{f}}{\bf{e}}_s^{xy},
\label{equ_network}
\end{equation}
where ${{\bf{Y}}_r} = {{\bf{Y}}_a} - {{\bf{Y}}_b}{\bf{Y}}^{-1}_d{{\bf{Y}}_c}$ after algebraic manipulations.
Additionally, a boundary condition should always be considered  according to Kirchhoff's current law, i.e., $\sum {i_{vsc}^x}  + \sum {i_{load}^x}  + i_g^x = 0$.

\begin{figure}[htbp]
	\centering
	\includegraphics[width=0.8\linewidth]{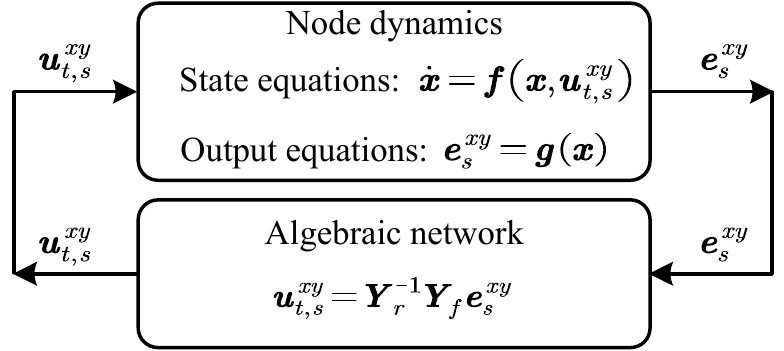}
	\caption{Schematic show for PEDPS.}
	\label{detailed}
\end{figure}

Therefore, although the network is still dynamic and described by differential equations, with the above analysis it can be well described by algebraic equations (\ref{equ_network}). In the meantime, the inductance dynamics between the internal and terminal voltages can be well incorporated into the node dynamics. Consequently, the whole system can be modeled by combining the node dynamics and algebraic network, which are described by nonlinear differential equations and linear algebraic equations, respectively. Accordingly, the internal and terminal voltages of (source) nodes are chosen as input-output ports between the node and network. The impact of dynamic network becomes clear, namely, it serves as a voltage divider and generates the terminal voltage based on the internal voltage of the sources instantaneously. The diagram is summarized in \figurename \ref{detailed}, clearly showing a similar structure to the traditional stability analysis \cite{kundur1994power}.

\section{Simulation Verification}

\begin{figure}[htbp]
	\centering
	\includegraphics[width=0.8\linewidth]{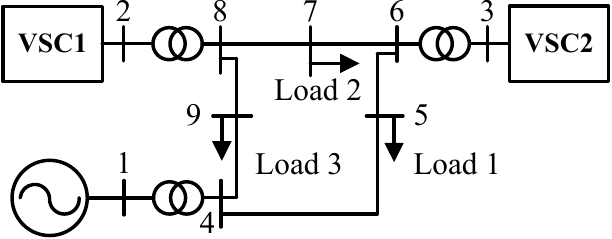}
	\caption{A modified 3-machine-9-bus system.}
	\label{9bus}
\end{figure}

\begin{figure}[htbp]
	\centering
	\includegraphics[width=0.8\linewidth]{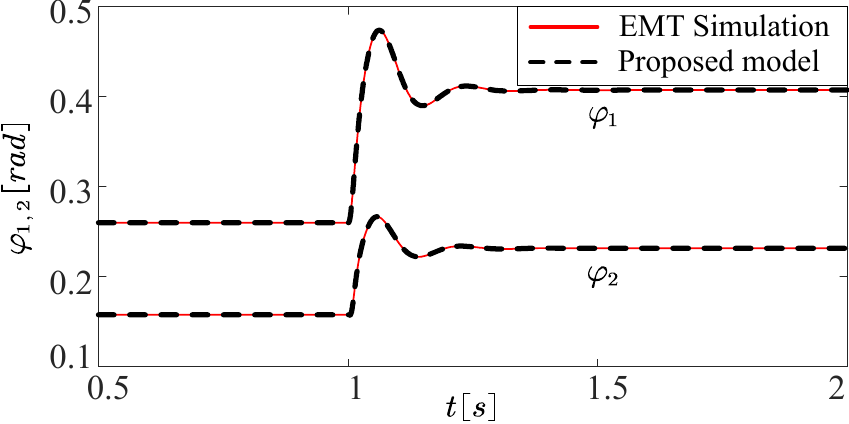}
	\caption{Comparison of time series between the EMT simulation (solid curves) and the proposed model (dashed curves) for the IEEE 9-bus system.}
	\label{time_verification}
\end{figure}

Broad time domain simulations have been conducted to verify the above established model. For example, a modified IEEE 3-machine-9-bus power system is tested, as shown in \figurename \ref{9bus}. Except for the slack bus 1, the SGs connected to buses 2 and 3 are replaced by VSCs. For convenience, the network parameters are unchanged except for the ground capacitance and line resistance neglected \cite{kundur1994power}. The details are given in Appendix.
When the active current reference of VSC1 increases from $1.59$pu to $2.0$pu, the time series of $\varphi_1$ and $\varphi_2$ for the phase differences between VSCs and slack node are illustrated in \figurename \ref{time_verification}, where solid and dashed curves represent the results of the EMT simulation and the proposed model, respectively. Clearly, as our model does not ignore any system information, they are identical.

Specifically, in the test, each VSC (nodes $2,3$) has two state variables for filter inductance and four state variables for controllers including the ACC and PLL. We can also consider more detailed controllers for node dynamics.
Each load (nodes $5,7,9$) has two state variables. The current injections of the infinite bus at node 1 are chosen as algebraic variables due to Kirchhoff's current law as above.
Meanwhile, all terminal voltages of source nodes, i.e., nodes $1,2,3,5,7,9$, are algebraic variables, while the terminal voltages of intermediate nodes, i.e., nodes $4,6,8$, are eliminated by (\ref{ini_net}). Therefore, there are 18 differential equations and 14 algebraic equations (including two boundary conditions).

\section{Conclusion}
In conclusion, a novel nonlinear model including node and network for the PEDPS has been well established, in which the node and network dynamics are described by nonlinear differential equations and linear algebraic equations, respectively. All independent state variables are kept, without ignoring any system information. The inductance dynamics has been well incorporated into the node dynamics, and in contrast, the original dynamic network which should be described by differential equations, has been well described by algebraic equations. With this differential-algebraic description, which most of electric power engineers are familiar with, the contributions of node and network and their relation also become apparent, i.e., the (source) nodes works as a dynamic source, whereas the network working as a voltage divider provides an instantaneous  rigid interaction. Clearly this model could contribute to an improved insight in the dynamics in PEDPS.

\appendix Parameters in the modified IEEE 3-machine-9-bus system are 
$f_{0} = 50$Hz, $L_f = 0.01$pu, $L_g = 0.01$pu.
Two VSCs adopt the identical ACC and PLL, and their parameters are 
$k_{p,acc} = 0.3$, $k_{i,acc} = 160$ for the ACC, and  $k_{p,pll} = 50$, $k_{i,pll} = 2000$ for the PLL.

\bibliographystyle{IEEEtran}
\bibliography{ref}

\end{document}